\documentclass[prd,twocolumn,nofootinbib, preprintnumbers, superscriptaddress]{revtex4-1} 

\usepackage{amsmath,amssymb,amscd,simplewick}
\usepackage{listings}
\usepackage{dsfont}
\usepackage{slashed}
\usepackage{color}
\usepackage{ulem}
\usepackage{float}
\usepackage{cancel}
\usepackage{graphicx}
\usepackage{epstopdf}
\usepackage{subfigure}
\usepackage{epsfig}
\usepackage{xcolor}
\usepackage[colorlinks=true,
            linkcolor=blue,
            urlcolor=blue,
            citecolor=green,          
            bookmarks=true,
            bookmarksnumbered=true,
            breaklinks=true,
            pdfpagemode=Fullscreen,
            pdfstartview=FitBH]{hyperref}

\hypersetup{pdfauthor = {Shao-Feng Ge},
	     pdftitle = {}, 
	     pdfsubject = {}, 
             pdfkeywords = {}, 
	     pdfcreator = {LaTeX with hyperref package},
	     pdfproducer = {dvips + ps2pdf} }

\definecolor{gesfpurple}{rgb}{0.47,0.19,0.42}

\definecolor{gesflanse}{rgb}{0.00,0.50,0.50}

\definecolor{gesfblue}{rgb}{0.08,0.42,0.76}

\definecolor{gesfred}{rgb}{1,0,0}

\definecolor{gesfwhite}{rgb}{1,1,1}

\definecolor{gesfblack}{rgb}{0,0,0}

\newcommand{\gsec}[1]{{\hypersetup{linkcolor=red}Sec.~\ref{#1}\hypersetup{linkcolor=blue}}}

\newcommand{\geqn}[1]{\hypersetup{linkcolor=blue}(\ref{#1})\hypersetup{linkcolor=blue}}
\newcommand{\gfig}[1]{{\hypersetup{linkcolor=violet}Fig.~\ref{#1}\hypersetup{linkcolor=blue}}}
\newcommand{\gtab}[1]{{\hypersetup{linkcolor=gesflanse}Tab.~\ref{#1}\hypersetup{linkcolor=blue}}}

\definecolor{Orange}{cmyk}{0,0.61,0.87,0}
\definecolor{JungleGreen}{cmyk}{0.99,0,0.52,0}
\definecolor{OliveGreen}{cmyk}{0.64,0,0.95,0.40}
\definecolor{Brown}{cmyk}{0,0.81,1,0.60}
\definecolor{RoyalBlue}{cmyk}{0.71,0.53,0,0.12}
\definecolor{Gray}{cmyk}{0,0,0,0.40}
\definecolor{LightPink}{cmyk}{0.0,0.25,0,0}
\definecolor{LLightPink}{cmyk}{0.0,0.10,0,0}
\definecolor{LightBlue}{cmyk}{0.25,0,0,0}
\definecolor{LightGray}{cmyk}{0,0,0,0.2}

\begin{document}

\title{Probing Light Mediators in the Radiative Emission of Neutrino Pair}
\author{Shao-Feng Ge}
\email{gesf@sjtu.edu.cn}
\affiliation{Tsung-Dao Lee Institute \& School of Physics and Astronomy, Shanghai Jiao Tong University, China}
\affiliation{Key Laboratory for Particle Astrophysics and Cosmology (MOE) \& Shanghai Key Laboratory for Particle Physics and Cosmology, Shanghai Jiao Tong University, Shanghai 200240, China}
\author{Pedro Pasquini}
\email{ppasquini@sjtu.edu.cn}
\affiliation{Tsung-Dao Lee Institute \& School of Physics and Astronomy, Shanghai Jiao Tong University, China}
\affiliation{Key Laboratory for Particle Astrophysics and Cosmology (MOE) \& Shanghai Key Laboratory for Particle Physics and Cosmology, Shanghai Jiao Tong University, Shanghai 200240, China}

\begin{abstract}
We propose a new possibility of using the coherently enhanced
neutrino pair emission to probe light-mediator
interactions between
electron and neutrinos. With typical momentum transfer at the
atomic $\mathcal O(1$\,eV) scale, this process is extremely sensitive
for the mediator mass range $\mathcal O(10^{-3} \sim 10^4$)\,eV.
The sensitivity on the product
of couplings with electron ($g^e$ or $y^e$) and neutrinos
($g^\nu$ or $y^\nu$) can touch down to
$|y^e y^\nu| < 10^{-9} \sim 10^{-19}$ for a scalar mediator
and $|g^e g^\nu| < 10^{-15} \sim 10^{-26}$ for a vector one,
with orders of improvement from the existing constraints.
\end{abstract}

\maketitle 

\section{Introduction}

Since the reactor mixing angle $\theta_r$ ($\equiv \theta_{13}$)
was measured by the Daya Bay \cite{DayaBay} and RENO \cite{RENO}
experiments in 2012, the neutrino study has entered a precision
era \cite{deSalas:2020pgw,Esteban:2020cvm,Capozzi:2021fjo}.
With the next-generation neutrino oscillation
experiments,
the neutrino mixing angles and the mass squared differences
would be precisely measured. For example, the solar mixing
angle $\theta_s$ ($\equiv \theta_{12}$) and $\Delta m^2_a$
($\equiv \Delta m^2_{31}$) at JUNO \cite{Li:2016txk,CaoNuTurn2012,Ge:2012wj,Ge:2015bfa,JUNO:2015zny} as well as
the atmospheric mixing angle $\theta_a$ ($\equiv \theta_{23}$)
at atmospheric \cite{Ge:2013zua,Ge:2013ffa,Chatterjee:2013qus,Choubey:2013xqa,Ghosh:2015ena}
and accelerator \cite{Agarwalla:2013ju,Ghosh:2015ena,Minakata:2013eoa,Ballett:2016daj,Coloma:2014kca,Choubey:2013xqa,Chatterjee:2013qus,Das:2016bwe,Singh:2017vvh} experiments.
The only exception is the Dirac CP phase and the neutrino
interactions. While the CP measurement is also approaching
discovery threshold with T2K \cite{T2K} and NO$\nu$A \cite{NOvA},
the neutrino interaction is still much less constrained
\cite{Farzan:2017xzy,Proceedings:2019qno}. Without precision measurement on the
neutrino interactions, especially those possiblility beyond
the Standard Model (SM), one cannot conclude that the
nonzero neutrino mass
is the only new physics in the neutrino sector. Most importantly,
without new interactions beyond the SM (BSM),
there is even no explanation
of the nonzero neutrino mass. It is natural to expect
neutrinos to participate BSM interactions
and urgent to find sensitive probes.

The BSM neutrino interactions can appear as non-standard
interactions (NSI) \cite{Farzan:2017xzy,Proceedings:2019qno} and
effective neutrino masses \cite{Ge:2018uhz, Venzor:2020ova}.
In addition, new interactions
can also lead to neutrino beamstrahlung \cite{Berryman:2018ogk},
neutrino trident production 
\cite{Altmannshofer:2014pba,Magill:2016hgc,Ge:2017poy,Magill:2017mps,Ballett:2018uuc,Shimomura:2020tmg},
modified spectrum from leptonic meson decays
\cite{Barger:1981vd, Lessa:2007up}, 
and neutrino decay \cite{Berezhiani:1991vk, Acker:1992eh, Dias:2005jm, Coloma:2017zpg, deGouvea:2019qre}.
The presence of new light particles have impact 
in the early Universe by generating 
extra degrees of freedom \cite{Babu:2019iml, Chigusa:2020bgq} or 
delaying the neutrino freeze-out \cite{Bell:2005dr,Cyr-Racine:2013jua}.
Despite these ways of probing neutrino
interactions, most of them involve large momentum transfer and hence cannot probe light mediators
much below $\mathcal O(1$\,MeV).
The only exception is the NSI from
the coherent scattering that involves zero momentum transfer
by definition and hence can probe very light mediators
\cite{Smirnov:2019cae,Babu:2019iml}.
However, NSI as matter effect is only sensitive to the ratio
between coupling and mass as a whole,
being unable to provide more
information about the mediator.

On the other hand, astrophysics and cosmology can give
very good constraints.
For example, big bang nuclesynthesis (BBN) may provide some 
constraints from the degree of freedom of 
relativistic species in early universe 
\cite{Babu:2019iml, Chigusa:2020bgq} in addition to the effective 
neutrino masses \cite{Venzor:2020ova}. 
And the most stringent bound comes from 
cooling of astrophysical sources 
\cite{Hardy:2016kme,Heurtier:2016otg} which are 
model dependent \cite{DeRocco:2020xdt}. Nevertheless, most
of these can only indirectly constrain the coupling with either
electron or neutrino, not both of them.
This highlights the importance of Earth-based 
experiments to test the possible BSM interactions that
neutrinos can experience.
\cite{Jaeckel:2006xm}.

The dark matter (DM) direct detection experiments
can also probe neutrinos and their interaction with electron
or nuclei. The recent observations of electron recoil at Xenon detectors
\cite{Aprile:2020tmw,PandaX-II:2020udv,PandaX-II:2021nsg}
have prompted several
possible explanations with light mediator between neutrino
and electron, including scalar
\cite{Boehm:2020ltd,Khan:2020vaf,Khan:2020csx,Gao:2020wfr,GPS}
and vector
\cite{Boehm:2020ltd,Khan:2020vaf,Khan:2020csx,Bally:2020yid,AristizabalSierra:2020edu,Lindner:2020kko,Karmakar:2020rbi,Amaral:2020tga,Chala:2020pbn,Ibe:2020dly,GPS} bosons,
eletromagnetic interactions 
\cite{Khan:2020vaf,Khan:2020csx,Amaral:2020tga,Miranda:2020kwy,Babu:2020ivd,Chala:2020pbn,Brdar:2020quo,AristizabalSierra:2020zod} with massless photon,
and sterile neutrino \cite{Ge:2020jfn,Shoemaker:2020kji,Chen:2021uuw,Shakeri:2020wvk}.
Due to the intrinsic energy threshold $\sim \mathcal
O(1$\,keV) of DM experiments, the typical momentum
transfer is around $\mathcal O(10$\,keV) for electron
recoil and $\mathcal O(1$\,MeV) for nuclei recoil.
Although the situation can be improved for electron
recoil when compared with the usual neutrino
experiments, the improvement is not significant enough.

In this work we present a new possibility of using
the {\it radiative emission of neutrino pair} (RENP)
\cite{Yoshimura:2006nd,Fukumi:2012rn,
Hiraki:2018jwu,Tashiro:2019ghs} to
probe the BSM interactions mediated by a light mediator.
The $\mathcal O($eV) energy scale of atomic transitions
are perfect for investigating the existence and properties
of light mediators. 
We start by briefly summarizing the SM calculation
of the neutrino pair emission process in
\gsec{sec:nu_pair_emission} for both Dirac
and Majorana neutrinos.
Then \gsec{sec:light_mediators} considers
the modifications to the spectral
function of the photon emission in the presence
of light vector/scalar mediators with mass in the range of
$\mathcal O(10^{-3} \sim 10^4$)\,eV.
The smallness of the mediator
mass implies a significant mo\-di\-fi\-ca\-tion and hence
unprecedented sensitivity as shown in
\gsec{sec:sensitivity_New_phys}. Finally, we summarize
the most prominent results and features in \gsec{sec:conclusion}.

\section{Neutrino Pair Emission}
\label{sec:nu_pair_emission}

\begin{figure}[t]
\centering
\includegraphics[scale = 0.56]{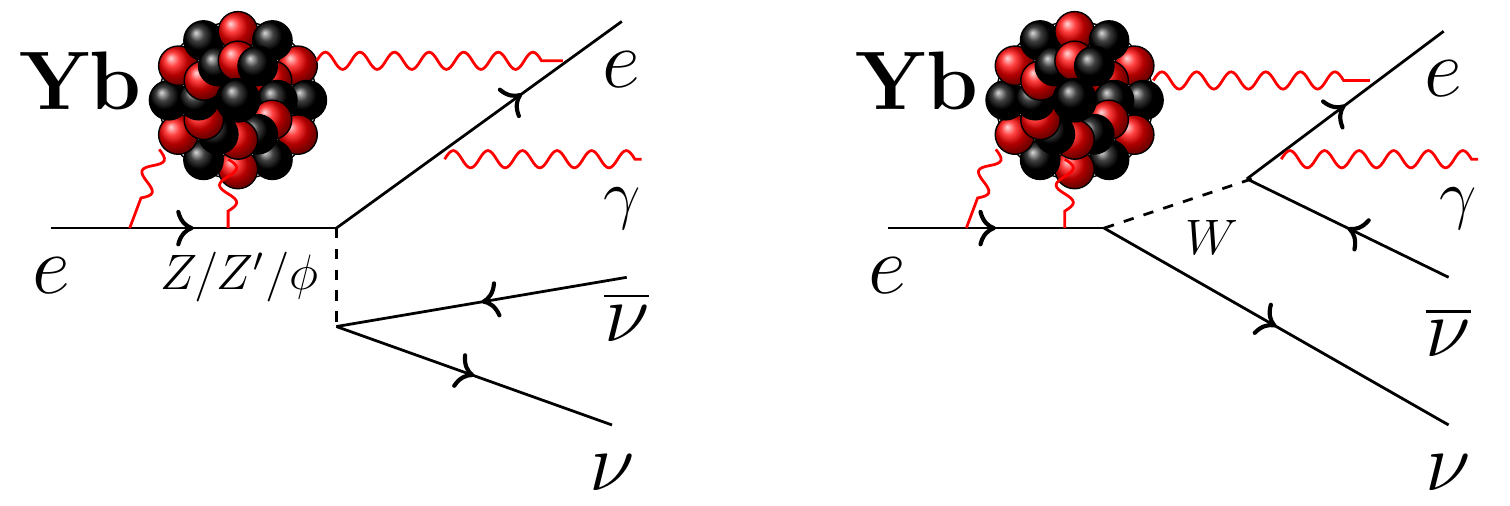}
\caption{Feynman diagrams for the radiative emission of neutrino pair with the SM heavy gauge bosons ($Z/W$) and light mediators
($Z'/\phi$).}
\label{fig:feynman_diagram}
\end{figure}

The {\it radiative emission of neutrino pair} (RENP)
\cite{Yoshimura:2006nd,Yoshimura:2011ri,Fukumi:2012rn,Hiraki:2018jwu,Tashiro:2019ghs}
is a coherently induced superradiance \cite{Yoshimura:2012tm}
of a neutrino pair and a photon when an
excited atomic state $|e\rangle$ transits to the
ground state $|g\rangle$,
$|e\rangle \rightarrow |g\rangle + \gamma + \nu \bar \nu$.
In the SM, the emission of neutrino
pair is mediated by the heavy $W/Z$ gauge bosons, as
illustrated in \gfig{fig:feynman_diagram}.
Since the gauge boson masses are much
larger than the typical atomic energy,
$m_W = 80.4\,\mbox{GeV}, m_Z = 91.2\,\mbox{GeV} \gg
\mathcal O(1 \sim 10)\,\mbox{eV}$,
a weak process at low energy 
is highly suppressed by the Fermi constant,
$G_{\rm F}\approx 1.67\times 10^{-5}$ GeV$^{-2}$.
Fortunately, the RENP can
be enhanced by two quantum mechanical effects
\cite{Yoshimura:2012tm}.
First, the bosonic nature of photon allows stimulated photon emission enhanced by a factor of the photon number density $n_\gamma$.
Second, the coherent behaviour of atoms can
further boost the decay rate by a factor of
$n_a^2$ where $n_a$ is the number of atoms
behaving coherently \cite{Dicke:1954zz}. For
macroscopically coherent
material, the enhancement can be large enough to
allow practical measurement.
 Since the emitted photon
spectrum can be precisely measured with electromagnetic instead of weak interactions, the RENP is especially sensitive to neutrino properties \cite{Yoshimura:2009wq,Dinh:2012qb},
including the neutrino mass
\cite{Song:2015xaa,Zhang:2016lqp},
the unitarity test of the neutrino mixing matrix
\cite{Huang:2019phr}, the Dirac/Majorana
nature of neutrinos \cite{Dinh:2012qb,Fukumi:2012rn},
the leptonic CP phases \cite{Dinh:2012qb,Yoshimura:2015yja},
the cosmic neutrino background \cite{Takahashi:2007ec,Yoshimura:2014hfa},
and light sterile neutrinos \cite{Dinh:2014toa}.
In the current section, we summarize the
key elements of RENP in the SM for both Dirac
and Majorana neutrinos.

\subsection{The SM Interactions for Atomic Transitions}
\label{sec:SM_prediction}

The RENP process involves both electromagnetic
and weak interactions,
\begin{equation}
  H = H_0 + D_\gamma + H_W.
\label{eq:total_hamiltonian_SM}
\end{equation}
The zeroth order $H_0$ describes the atomic energy
levels of electrons in an atom,
namely, $H_0|a\rangle = E_a |a\rangle$,
for the excited ($a = e$), virtual ($a = v$),
and ground ($a = g$) states, respectively.
For comparison, $D_\gamma$ represents the electric
dipole interaction while $H_W$ is the weak
counterpart.

The excited state $|e\rangle$ should be meta-stable
with $E_v > E_e > E_g$ such that the
transition between $| e \rangle$ and $| g \rangle$
cannot happen by emitting a single photon.
Otherwise, the excited state $| e \rangle$ would
decay much faster ($\sim1$\,ns \cite{Fukumi:2012rn})
to the ground state directly via $|e
\rangle\rightarrow |g\rangle+\gamma$ than the
RENP. This can be achieved by selecting the atom
with appropriate $| e \rangle$ and $| g \rangle$.
For example, the Xe (Yb) element has angular
momentum $J_e = 2$ and $J_g = 0$ ($J_e = 0$ and
$J_g = 0$) \cite{Song:2015xaa} for the excited and
ground states, respectively. With emission of a
single photon that carries only $J_\gamma = 1$, the
angular momentum can not conserve for the $J_e
\rightarrow J_g$ transition. This significantly
reduces the background that cannot be
observationally distinguished from the RENP signal
$|e\rangle \rightarrow |g\rangle + \gamma + \nu
\bar \nu$.

With the direct transition from $| e \rangle$ to
$| g \rangle$ forbidden, the neutrino pair
emission is at most a second order process through
a virtual atomic state $|v\rangle$ at higher
energy, $| e \rangle\rightarrow | v \rangle
\rightarrow | g \rangle$. The whole transition is
then a combined transition of the E1$\times$M1
type. The current experimental configuration
\cite{Huang:2019phr,Hiraki:2018jwu} allows an M1
transition for $| e \rangle\rightarrow | v \rangle
+ \nu \bar \nu$ with neutrino pair emission and an
E1 transition for $|v \rangle \rightarrow|g\rangle
+ \gamma$ with photon emission. Since $E_v > E_e$,
the first step $|e\rangle \rightarrow |v\rangle$
cannot happen separately. The two transitions have
to happen simultaneously.

For the electromagnetic part, the dominant
contribution comes from an electric dipole
transition (E1)  induced by $D_\gamma$. More
explicitly, the electric dipole is linearly
related to the laser electric field ${\bf E}({\bf
x})$ with trigger frequency $\omega$,
\begin{eqnarray}
   \langle g|D_\gamma| v \rangle 
\equiv
  \mathcal M_D  e^{-i\omega t + {\bf k}\cdot {\bf x}},
\quad
  \mathcal M_D
\equiv
- {\bf d}_{gv}\cdot {\bf E}_0,
\label{eq:relations_atomic}
\end{eqnarray}
where $\omega$ and ${\bf k}$ are the photon energy
and momentum, respectively. The 3-vector ${\bf
d}_{gv}$ sandwiched between the virtual and ground
states is the dipole operator for the $v\rightarrow g$
transition and ${\bf E}_0$
is the stimulating electric field. 

The $H_W$ term contains the SM weak interactions between electron and neutrinos,
\begin{eqnarray}
  \mathcal H_W
\equiv
  \sqrt 2 G_F
  \left( v_{ij} J^\mu_V - a_{ij} J^\mu_A \right)
  \bar \nu_{i L} \gamma_\mu \nu_{j L}.
\label{eq:weakH}
\end{eqnarray}
The electron current
$J_e^\mu \equiv v_{ij} J^\mu_V - a_{ij} J_A^\mu $
is a linear combination of the vector current
$J^\mu_V \equiv \langle v|\overline e \gamma^\mu e|e\rangle$
and the axial current
$J^\mu_A \equiv \langle v|\overline e \gamma^\mu \gamma_5 e|e\rangle$. The coefficients
\begin{eqnarray}
  a_{ij}
\equiv
  U_{ei} U^*_{ej}
- \frac{\delta_{ij}} 2,
~ {\rm and} ~
  v_{ij}
\equiv
  a_{ij}
+ 2 s^2_w \delta_{ij},
\end{eqnarray}
contain the PMNS matrix $U$ of neutrino mixing which is
a function of mixing angles and leptonic CP phases 
\cite{Zyla:2020zbs}.
From the weak Hamiltonian \geqn{eq:weakH},
we can obtain the matrix element for the 
$|e\rangle \rightarrow |v\rangle$ transition,
\begin{subequations}
\begin{eqnarray}
  \langle v | H_W|e\rangle
& \equiv &
  \mathcal M_W^{(D)} e^{-i(p_\nu+p_{\bar \nu})\cdot x}
\\ 
  \mathcal M_W^{(D)}
& \equiv &
  \sqrt{2}G_{\rm F} (J_e^\mu)_{ij}
  \bar u(p_{\nu_i})\gamma_\mu P_L v(p_{\bar\nu_j}),
\label{eq:Mw-Dirac}
\end{eqnarray}  
\label{eq:relations_atomic2}
\end{subequations}
for Dirac neutrinos.
The M1 transition is mainly contributed by the 
axial current that has only spatial components
and is proportional to the electron spin in
the nonrelativistic limit, $J_A^\mu \approx (0,\langle v| 2
{\bf S} |e\rangle)$ \cite{Song:2015xaa}. For comparison,
the electric dipole ${\bf d}_{gv}$ is parity odd (E1)
while the spin operator ${\bf S}$ is parity even (M1).

\subsection{Dirac vs Majorana Neutrino Matrix Elements}

The matrix element $\mathcal M^{(D)}_W$
\geqn{eq:Mw-Dirac} for Dirac neutrinos has only one
term, with $\bar u(p_{\nu_i})$ for neutrino and
$v(p_{\bar \nu_j})$ for anti-neutrino in the final
state. In comparison, Majorana neutrinos can
contribute an extra term since there is no
distinction between neutrino and anti-neutrino now.
In the formalism of second quantization, the
Majorana neutrino field, $\nu \sim a u + a^\dagger
v$, has only one set of annihilation ($a$) and
creation ($a^\dagger$) operators. Both $\bar \nu$
and $\nu$ fields in \geqn{eq:weakH} can
create/annihilate neutrino and anti-neutrino final
states \cite{Dinh:2012qb},
\begin{eqnarray}\nonumber 
  \mathcal M_W^{(M)} 
=
  \sqrt{2} G_F 
  J_A^\mu 
  \Bigl[ 
& &
  a_{ij} 
  \bar u
  (p_{\nu_i}) 
    \gamma_\mu P_L 
  v (p_{\overline{\nu}_j})
\\
- && 
  a_{ij}^* 
  \bar u(p_{\overline \nu_j})
    \gamma_\mu P_L 
  v (p_{\nu_i})
  \Bigr].
\label{eq:Mw-Majorana}
\end{eqnarray}
The minus sign comes from the commutation property
of fermion fields and we have used the property
$a_{ji} = a_{ij}^*$ to write the coefficients all
in terms of $a_{ij}$.
Using the spinor relations $\overline v = -  u^T C$ and
$u = - C \overline v^T$ and $C^{-1} \left(\gamma_\mu P_L \right)^T C =
 - \gamma_\mu P_R$,
where $C$ is the charge conjugation operator,
the squared matrix elements become,
\begin{subequations}
\begin{eqnarray}
  |\mathcal M_W^{(D)}|^2
& = & \nonumber 
  2 G_F^2 J_A^\mu J_A^{\nu *} 
  |a_{ij}|^2
\\ & \times &
    \bar u(p_{\nu_i})
    \gamma_\mu P_L 
  v (p_{\overline \nu_j})
    \bar v (p_{\overline \nu_j})
    \gamma_\nu P_L 
  u(p_{\nu_i}),
\\
  |\mathcal M_W^{(M)}|^2
& = &
  |\mathcal M_W^{(D)}|^2
-
  2 G_F^2
  {\rm Re}
  \Bigl[ 
    a_{ij}^2
  J_A^\mu J_A^{\nu *}
\nonumber
\\ & & 
    \bar u(p_{\nu_i})
      \gamma_\mu P_L 
    v(p_{\overline{\nu}_j})
    \bar v(p_{\overline{\nu}_j})
      \gamma_\nu P_R
   u(p_{\nu_i})
  \Bigr].
\quad
\end{eqnarray}
\end{subequations}
We have also included a factor of $1/2$ in
$|\mathcal M_W^{(M)}|^2$ to avoid double counting 
in the phase space integration. 
The appearance of both chirality projection operators
$P_L$ and $P_R$ will result in chirality flip and hence
the extra term is proportional to a product of both
neutrino masses $m_i m_j$.
Since the final-state momentum of neutrinos from atomic
transition are typically $\mathcal O(1$eV),
this chirality-flipping term is no longer suppressed.
This is the reason why RENP can be sensitive to the
neutrino masses and hence its Dirac/Majorana nature.

The squared electroweak matrix element after averaging over
spins and all the possible atomic magnetic numbers is, 
\begin{eqnarray}
  \overline{|\mathcal M_W^{ij}|^2} 
& = &
 \frac 8 3(2 J_v + 1) |a_{ij}|^2 G_{\rm F}^2 C_{ev}
\label{eq:MW_average}
\\
& \times & 
\Bigl\{
  2 {\bf p}_{\nu_i}\cdot {\bf p}_{\bar \nu_j}
+ 3 p_{\nu_i} \cdot p_{\bar \nu_j}
- 3 \delta_M {\rm Re}[a_{ij}^2] m_i m_j
\Bigr\}.
\nonumber
\end{eqnarray}
The prefactor  $\delta_M = 1$ for Majorana
neutrinos and $\delta_M = 0$ for Dirac ones. To
obtain \geqn{eq:MW_average}, the electron spins
have also been summed over \cite{Dinh:2012qb},
\begin{equation}
  \frac{1}{(2 J_e + 1)}\sum_{m_e m_v}
  \langle v|S^i |e\rangle \langle e| S^j | v\rangle 
=
  (2 J_v + 1) \frac{C_{ev}}{3} \delta^{ij},
  \label{eq:Sij}
\end{equation}
where $J_e$ and $J_v$ are the total spins of the excited
($|e\rangle$) and the virual ($|v\rangle$) states,
respectively. The constant $C_{ev}$ is a factor that 
depends on the atom. For Yb we have $(2 J_v + 1)
C_{ev} = 2$ \cite{Song:2015xaa}.

\subsection{Emission Rate and Spectrum}

The photon emission rate of the reaction is~\cite{Dinh:2012qb,Song:2015xaa,Zhang:2016lqp},
\begin{eqnarray}\label{eq:diff_xsec}
  d\Gamma_{ij}
=
  \frac{\Gamma_0}
  {(E_{vg} - \omega)^2 \omega}
  \frac
  {\overline{|\mathcal{M}^{ij}_W|^2}}
  {8 G_F^2 C_{ev} (2J_v + 1)}
  dE_\nu,
\end{eqnarray}
with the reference decay width $\Gamma_0$
\cite{Song:2015xaa,Zhang:2016lqp},
\begin{subequations}
\begin{eqnarray}
  \Gamma_0 
& \equiv &
  (2J_v + 1)\frac{n_a^2 C_{ev} G_{\rm F}^2 |{\bf d}_{gv}\cdot {\bf E}_ 0|^2}{\pi }
\\ & \approx &
  0.002{\rm s}^{-1}
  \left(\frac{V}{10^2 {\rm cm}^3}\right)
  \left(\frac{n_a^2 n_\gamma }{10^{21}{\rm cm}^{-3}}\right)
  \eta(t).
\end{eqnarray}
\end{subequations}
For illustration, we have used Yb. The decay width
scales linearly with the target volume $V$ and the
fraction $\eta$ of atoms that behave
coherently. An $\eta \approx 1$ is claimed to be
achievable \cite{BoyeroGarcia:2015dye} and we set
$\eta = 1$ for simplicity.

The photon kinematics of the
RENP process is fixed by the two back-to-back
trigger laser beams whose frequencies
\cite{Song:2015xaa,Zhang:2016lqp} ($\omega_1$ and
$\omega_2$) sum up to $\omega_1 + \omega_2 = E_e -
E_g$. The stimulated emission imposes a fixed
emitted laser frequency $\omega = \omega_1
<\omega_2$ in the same direction of the $\omega_1$
laser beam and also enforces the energy/momentum
conservation,
\begin{eqnarray}
  E_e - E_g = \omega + E_\nu + E_{\overline \nu}
\quad {\rm and} \quad 
{\bf k} = - {\bf p}_\nu - {\bf p}_{\overline \nu}\,.
\label{eq:kinematic_conditions}
\end{eqnarray}
The invariant mass of the neutrino pair is then related to
the photon energy,
$m^2_{\nu \bar \nu} = E^2_{eg} - 2 E_{eg} \omega \geq (m_i + m_j)^2$,
which imposes a frequency upper limit on the emitted photon,
\begin{eqnarray}
  \omega_{ij}^{\rm max}
\equiv
  \frac{E_e - E_g}{2} 
- \frac 1 2\frac{(m_i + m_j)^2}{(E_e - E_g)}.
\label{eq:maximum_frequencies_thresholds}
\end{eqnarray}
Only for $\omega < \omega_{ij}^{\rm max}$, the RENP
process can be triggered.

With the emitted photon frequency $\omega$ fixed,
the 3-body final state only needs 2-body phase
space integration which can reduce to a single
integration \geqn{eq:diff_xsec} over the neutrino
energy $E_\nu$. The integration limit is
a function of the photon frequency $\omega$,
\begin{equation}
  \bar E - \frac{\omega \Delta_{ij}(\omega)} 2
\leq E_\nu \leq
  \bar E + \frac{\omega \Delta_{ij}(\omega)} 2,
\end{equation}
where,
\begin{subequations}
 \begin{eqnarray}\nonumber
    \Delta_{ij}(\omega)
& \equiv &
    \frac{ \sqrt{E_{eg}(E_{eg}-2 \omega)-(m_i+m_j)^2} }{E_{eg} (E_{eg}-2 \omega)}
\\&\times&
      \sqrt{E_{eg}(E_{eg}-2 \omega)-(m_i-m_j)^2},
\end{eqnarray}
\begin{eqnarray}
  \bar E
\equiv
  \frac {[E_{eg}(E_{eg}-2\omega) -  \Delta m^2_{ij}](E_{eg}-\omega)}
        {2E_{eg}(E_{eg}-2\omega)}.
\end{eqnarray}
\end{subequations}
 
After integration over the kinematic
range in \geqn{eq:kinematic_conditions} 
and sum over all the neutrino mass
eigenstates, the total decay width becomes
$\Gamma \equiv \Gamma_0 \mathcal I(\omega)$
\cite{Dinh:2012qb}.
The spectral function $\mathcal I(\omega)$
is defined as,
\begin{eqnarray}
  \mathcal I(\omega)
\equiv \nonumber 
  \sum_{ij} 
  & &
  \frac{\Delta_{ij}(\omega) 
  }{(E_{vg} - \omega)^2} \Theta(\omega - \omega_{ij}^{\rm max})
\\ &  &
\left[
  |a_{ij}|^2 I_{ij}^{(D)} 
-
  \delta_M {\rm Re}[a_{ij}^2] I_{ij}^{(M)}
\right],
\label{eq:SM_spectral_function}
\end{eqnarray}
with a Heaviside theta function $\Theta$  to impose frequency requirement and,
\begin{subequations}\label{eq:I_Dirac_and_Maj}
\begin{eqnarray}
  I_{ij}^{(D)}
& \equiv &
 \frac{1}{3}
\left\{
  E_{eg}
  (E_{eg}-2\omega)
-
  \frac{1}{2}(m_i^2+m_j^2)
\right.
\\ &+ &\nonumber
\left.
  \frac{\omega^2}{2}
  \left[ 1-\frac{1}{3}\Delta_{ij}(\omega)^2\right]
-
  \frac{(E_{eg}-\omega)^2(\Delta m_{ij}^2)^2}{2E_{eg}^2(E_{eg}-2 \omega)^2}
\right\},
\\ 
I_{ij}^{(M)}
& \equiv &
m_i m_j\,.
\end{eqnarray}
\end{subequations}

There are two candidate elements that allow 
RENP process: Xenon (Xe) and Ytterbium (Yb)
\cite{Dinh:2012qb}. 
In this work we focus on Yb that 
can reach $\mathcal O(20)$ events per day for a
target volume of $100$ cm$^{3}$ and atomic
density  of $10^{21}$ cm$^{-3}$
\cite{Zhang:2016lqp}.
Although these parameters are not achievable yet,
there have been some advances recently \cite{Hiraki:2018jwu}.

\begin{figure}[t]
  \centering
  \includegraphics[scale = 0.22]{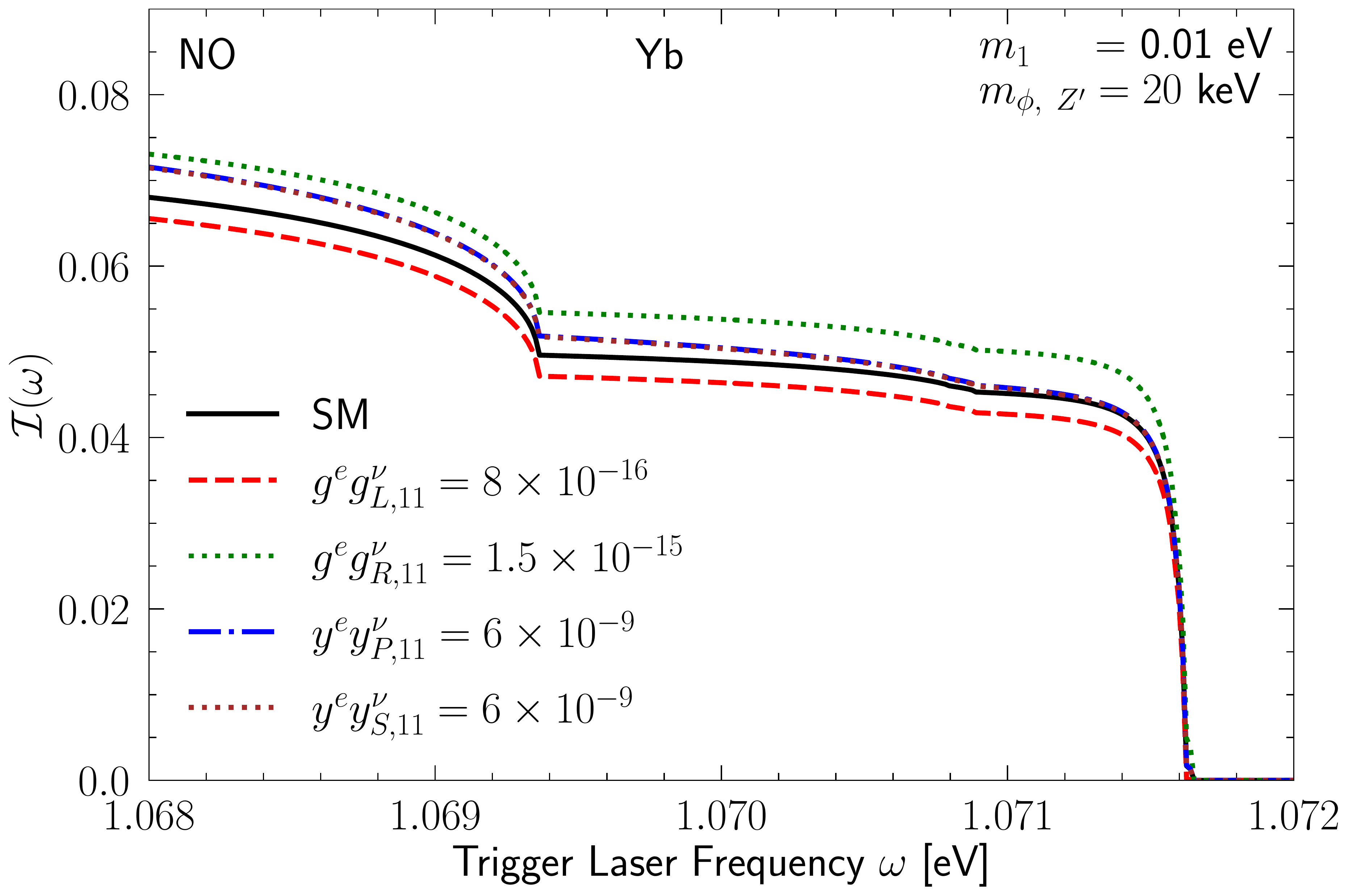}
\caption{
The spectral function $\mathcal I(\omega)$ of Yb as a function of 
the trigger laser frequency $\omega$ for the normal ordering (NO)
and lightest neutrino mass $m_1 =
0.01$ eV in the SM (black) as well as in the presence 
of a light mediator with mass $m_{\phi,
Z'} = 20$ keV (colorful lines).
For comparison, we show four different
scenarios of the BSM couplings:
$g^e g^\nu_{L,11} = 8\times 10^{-16}$ (red dashed)
$g^e g^\nu_{R, 11} = 1.5\times 10^{-15}$ (green dotter)
$y^e y^\nu_{S, 11} = 6\times 10^{-9}$ (brown dotted)
$y^e y^\nu_{P, 11} = 6\times 10^{-9}$ (blue dot-dashed).
Notice that the scalar and pseudo-scalar
contributions almost overlap with each other.}
\label{fig:new_spectral_function}
\end{figure}

The black line of \gfig{fig:new_spectral_function} 
illustrates the SM spectral function 
\geqn{eq:SM_spectral_function} as a function of the
trigger laser frequency $\omega$ for the normal
neutrino mass ordering (NO) and li\-gh\-test 
neutrino mass $m_1 = 0.01$ eV
with Dirac neutrino. 
We can see several kinks that correspond to the 6
kinematic upper limits in 
\geqn{eq:maximum_frequencies_thresholds}.
The two most prominent kinks are 
$\omega_{33} \approx 1.0693$ eV and $\omega_{13} 
\approx 1.0710$ eV. The curve drops to 
zero at the highest upper limit
$\omega^{\rm max}_{11} \approx 1.0716$ eV.

\section{Light Mediators}
\label{sec:light_mediators}

For the SM interactions discussed above,
the $W/Z$ gauge bosons essentially contribute
as contact operators, $1 / (q^2 - m^2) \approx
1/m^2$ with $m^2 \gg q^2$. For atomic transitions, the momentum transfer is
typically eV scale, $q^2 \sim \mathcal
O(\mbox{eV}^2)$, in contrast to $m^2_W, m^2_Z 
\sim \mathcal O(10^{21} \mbox{eV}^2)$.
The hierarchy between the cut-off scale and the
physical energy is as large as 21 orders at the
amplitude level and more than 40 orders for the decay
width. If the
mediator mass is lowered down to the atomic scale,
the decay width can be significantly enhanced.
In other words, the RENP is very sensitive to
probing the coupling
of light mediators with electron and neutrinos.

Generally speaking, a light mediator between
electron and neutrinos can be either scalar or
vector boson. In principle, both charged and
neutral mediators are allowed. However, a charged
light mediator with mass at eV scale can be
easily detected and hence has probably been
excluded already. We only consider neutral light
mediators for illustration.

A scalar boson can have both scalar and
pseudo-scalar couplings with fermion while a
vector boson can mediate both vector and
axial-vector interactions. However, the $|e\rangle
\rightarrow |v\rangle + \nu \bar \nu$
transition only allows M1 transition for the
coupling with electron. Not all
interactions can contribute. In
\gtab{tab:processes_types} we summarize all the
possible Lorentz structures for new interactions
with electron in the non-relativistic limit.
Only the pseudo-scalar and
axial-vector interactions are M1 type.
Nevertheless, the neutrino part can still
have all possible interaction types including
the vector and scalar ones.

\begin{table}[b]
\centering
\begin{tabular}{c|c|c|c}
     Coupling    & $\mathcal L_{\rm new}$ &  Non-Relativistic Transition   &  Type\\
     \hline
    scalar           & $y^e_S\bar e e $    & $\langle f |i\rangle$ & E1 \\
    pseudo-scalar    & $iy^e_P\bar e \gamma_5 e$    & $\frac{{\bf q}}{2m_e}\cdot \langle f | \boldsymbol \sigma |i\rangle$ & M1 \\
    vector    & $g^e_V\bar e \gamma^\mu e $    & $(\langle f |i\rangle, \frac{{\bf q}}{2m_e}\cdot \langle f | \boldsymbol \sigma \boldsymbol \sigma  |i\rangle$) & E1 \\
    axial-vector    & $g^e_A\bar e \gamma^\mu \gamma_\mu e $    & $(\frac{{\bf q}}{2m_e}\cdot \langle f | \boldsymbol \sigma |i\rangle, \langle f |\boldsymbol \sigma |i\rangle)$ & M1
\end{tabular}
\caption{The electron atomic transition currents from an
initial state $|i \rangle$ to a final state $|f\rangle$ 
for non-relativistic electrons and momentum transfer ${\bf q}$ 
\cite{Weinberg:1995}.}
\label{tab:processes_types}
\end{table}

Light mediators with masses below $\mathcal O(10)$ keV that couple to
neutrinos can appear in many models. An incomplete list
includes $U(1)$ extensions of the SM gauge group
\cite{Gelmini:1980re, Chikashige:1980qk,Machado:2010ui,Machado:2013oza,Fayet:2016nyc,Heeck:2010pg}, 
fifth force \cite{Wise:2018rnb,Dror:2020fbh,Bustamante:2018mzu},
$U(1)_R$ symmetry connected to a hidden right-handed neutrino sector \cite{Alikhanov:2021dhb,Xu:2020qek},
neutrino mass models \cite{Bonilla:2016zef,Shakya:2018qzg,Chauhan:2020mgv},
the neutrinophilic
two-Higgs doublet models \cite{Machado:2015sha,Lindner:2020kko},
the neutrinophilic axion-like particles \cite{Mohapatra:1982tc,Arason:1990sg,Huang:2018cwo,Baek:2019wdn,Baek:2020ovw}, 
neutrino non-standard interaction with light 
mediators \cite{Smirnov:2019cae,Coloma:2020gfv}.
Also, light mediators can be a connection between neutrinos and ultra-light dark matter
\cite{Berlin:2016woy,Brdar:2017kbt,Farzan:2019yvo,Cline:2019seo,Ge:2019tdi,Choi:2019zxy,Dev:2020kgz,Alonso-Alvarez:2021pgy} and inflation \cite{Joshipura:2020ibd}.
Some discussions on light mediator and neutrino  interactions
can also be found in \cite{He:1998ng,Ding:2020yen}.
The rich phenomenology of light mediators in the leptonic sector
can be found in \cite{Babu:2019iml}. It is well motivated to find possible ways of probing light and ultra-light mediators using low enough momentum transfer as we elaborate below.

\subsection{Vector Mediator}

Since the vector coupling with electron does not
contribute to the RENP process, the relevant
new interactions with a vector mediator $Z'$ are,
\begin{eqnarray}
  \mathcal L_V
=
  g^e \bar e \gamma^\mu \gamma_5 e Z'_\mu 
+ \bar \nu_i \gamma^\mu (g^\nu_{L,ij} P_L + g^\nu_{R,ij} P_R) \nu_j Z'_\mu.
\qquad
\end{eqnarray}
Extra terms can arise as correction to the scattering matrix elements \geqn{eq:Mw-Dirac} and \geqn{eq:Mw-Majorana},
$\mathcal M_W^{ij} \rightarrow \mathcal M_{\rm tot}^{ij} \equiv \mathcal M_W^{ij} + \mathcal M_{\rm New}^{ij}$, 
where $\mathcal M_{\rm New}^{ij}$ is the new contribution,
\begin{eqnarray} \nonumber 
  \mathcal M_{\rm tot}^{ij}
=
&&
  \sqrt 2
  G_{\rm F} J^\mu_A
  \Bigl[
    \bar u(p_{\nu_i}) \gamma_\mu
    \left( a^L_{ij} P_L + a^R_{ij} P_R \right)
    v (p_{\bar \nu_j})
\\ &&
- 
   \delta_M
    \bar u(p_{\bar \nu_j}) \gamma_\mu
    \left( a^L_{ji} P_L + a^R_{ji} P_R \right)
    v (p_{\nu_i})
  \Bigr].
\label{eq:M-vector}
\end{eqnarray}
Although only the left-handed neutrino current is
involved in the SM, the new interactions allow
the right-handed component to also participate,
\begin{subequations}
\begin{eqnarray}
  a^L_{ij}
& \equiv &
  a_{ij}
+
  \frac {g^e g^\nu_{L,ij}}
		  	{\sqrt 2 G_F (q^2 - m^2_{Z'})}
  \delta_{ij},
 \label{eq:aL_parameter}
\\
  a^R_{ij}
& \equiv &
  \frac {g^e g^\nu_{R,ij}}
				{\sqrt 2 G_F (q^2 - m^2_{Z'})} \delta_{ij}.
\end{eqnarray}  
\end{subequations}
The first term $a_{ij}$ is the SM contribution and 
those terms involving $1/(q^2 - m^2_{Z'})$ come
from the $Z'$ propagator. Since an overall Fermi
constant $G_F$ has been extracted in \geqn{eq:M-vector},
the new terms also contain a $G_F$ in the denominator
although the weak scale is not necessarily relevant
here.

The total matrix element
squared $\overline{|\mathcal M^{ij}_{\rm tot}|^2}$
after averaging over all spin and atomic state is,
\begin{eqnarray}
&&
  \frac 8 3 (2 J_v + 1) C_{ev} G^2_F
\\ & & \times \nonumber 
\left[
  \Bigl(|a^L_{ij}|^2 + |a^R_{ij}|^2 
  - 
  2 \delta_M {\rm Re}[a^L_{ij}a^R_{ij}]
  \Bigr)
  (2 {\bf p}_i \cdot {\bf p}_j + 3 p_i \cdot p_j)
\right.
\\ & & 
\left.
\hspace{4mm}
+ 
 3 m_i m_j
  \left(
     2 {\rm Re}[a^{L *}_{ij} a^R_{ij}]
  - \delta_M {\rm Re}[(a^L_{ij})^2 + (a^R_{ij})^2]
  \right)
\right].
\nonumber 
\end{eqnarray}
It is interesting to observe that the first
line in the bracket has the same structure as
\geqn{eq:MW_average} with the overall factor
substitution $|a_{ij}|^2 \rightarrow |a^L_{ij}|^2
+ |a^R_{ij}|^2 - 2 \delta_M {\rm Re} [ a^L_{ij}
a^R_{ij}]$. The last term in the bracket comes
from the interference of the right- and
left-handed components and is proportional to the
neutrino masses. It is also similar to the
Majorana contribution in the SM with the
replacement, ${\rm Re}[a_{ij}^2] \rightarrow 2
{\rm Re}[a^L_{ij}a^R_{ij}] - \delta_M {\rm Re}
[(a^{L *}_{ij})^2 + (a^R_{ij})^2]$.

Correspondingly, the spectral function $\mathcal I(\omega)$ in
\geqn{eq:SM_spectral_function} receives the following
correction,
\begin{eqnarray}
	\mathcal I_{Z'}
& \equiv &
  \sum_{ij} 
  \frac{\Delta_{ij}(\omega) 
  }{(E_{vg} - \omega)^2} 
  \Theta(\omega - \omega_{ij}^{\rm max})
\nonumber 
\\
& \times &
\left[
  \Bigl(
	|a^L_{ij}|^2 + |a^R_{ij}|^2 
  - 2 \delta_M {\rm Re} [a^L_{ij} a^R_{ij}]
  \Bigr)
  I_{ij}^{(D)} 
\right.
\nonumber
\\
& &
\hspace{-2mm}
+
\left.
  \Bigl(
       \delta_M
    {\rm Re} \left[ (a^L_{ij})^2 + (a^R_{ij})^2 \right]
   -
    2{\rm Re} \left[ a^{L*}_{ij}a^R_{ij} \right]
   \Bigr)
    I_{ij}^{(M)}
\right].
\qquad
\label{eq:Iz}
\end{eqnarray}
\gfig{fig:new_spectral_function} shows the 
changes in the spectral function $\mathcal I$ 
for a light mediator mass of $m_{Z'} = 20$\,keV
and  two different coupling combinations (1) $g^e
g^\nu_{L,11} = 8\times 10^{-16}$ (red dashed) and
(2) $g^e g^\nu_{R,11} = 1.5\times 10^{-15}$
(dotted green). As we can see, the $g_L^\nu$
coupling can be smaller than $g_R^\nu$ and still
produce a comparable modification of the spectral
function. This is because of the interference 
between the SM coupling $a_{ij}$ and the new 
couplings as squared terms of $a^L_{ij}$.
This interference with the large SM coupling
$a_{ij}$ only applies to $g^e g^L_{ij}$ but not
$g^e g^R_{ij}$. In addition, there is an
interference term between $a_{ij}^L$ and
$a_{ij}^R$ due to the chirality flip. For
the Dirac case it is proportional to the neutrino
masses and hence is a small correction. In
comparison, the interference term for Majorana
neutrinos is much larger with its coefficient
being momentum instead of the tiny neutrino
masses.

\subsection{Scalar Mediator}

As explained above, only pseudo-scalar interaction
on the electron side can contribute the required
M1 transition while the neutrino side is general,
\begin{eqnarray}
  \mathcal L_S
& \equiv &
  i y^e_P \bar e \gamma_5 e \phi
+  \bar \nu_i 
  (y^\nu_{S,ij} + i \gamma_5 y^\nu_{P,ij})
  \nu_j \phi
+ h.c.,
\quad
\end{eqnarray}
with both scalar ($y^\nu_{S,ij}$) and pseudo-scalar
($y^\nu_{P,ij}$) interactions.
The corresponding new matrix element is,
\begin{eqnarray}
  \mathcal M^{ij}_S
& = & 
  \frac {
    y^e J_P \bar u(p_{\nu_i}) 
    (y^\nu_{S,ij} + i \gamma_5 y^\nu_{P,ij})
    v(p_{\overline \nu_j})
  }{E_{eg} 
  (E_{eg} - 2 \omega) - m^2_\phi}
\nonumber 
\\
& - & 
  \delta_M
  \frac {
    y^e J_P
    \bar u(p_{\overline \nu_j}) 
    (y^\nu_{S,ji} - i \gamma_5 y^\nu_{P,ji})
    v(p_{\nu_i})
  }{E_{eg} 
  (E_{eg} - 2 \omega) - m^2_\phi},
\qquad
\end{eqnarray}
where the second line arises only for Majorana neutrinos.

The total averaged matrix element squared,
\begin{eqnarray}
  \overline{|\mathcal M_{\rm tot}^{ij}|^2}
=
  \overline{|\mathcal M^{ij}_W|^2}
+ \overline{|\mathcal M^{ij}_S|^2}
+ 2 {\rm Re} 
  \left( \overline{ \mathcal M^{ij}_W \mathcal M^{ij *}_S} \right),
\qquad
\end{eqnarray}
receives corrections $\overline{|\mathcal M^{ij}_S|^2}$
from the scalar mediator and the $W,Z/\phi$ interference term. Note that the matrix
element $\mathcal M^{ij}_W$ contributed by the weak
interactions contains an axial-vector current
$J^\nu_A$ of electron while the scalar one $\mathcal M^{ij}_S$
by the light scalar mediator contains a pseudo-scalar
counterpart $J_P \equiv \frac 1 {2 m_e}
{\bf q} \cdot \langle e | \boldsymbol \sigma | v \rangle$ as summarized in \gtab{tab:processes_types}. Then, the
square of the new matrix element contains a term
$|J_A|^2$ and the interference one $J_A^\mu
J^*_{P}$. Using the spin relation \geqn{eq:Sij} we
can obtain,
\begin{subequations}
\begin{eqnarray}
  \frac{1}{(2 J_e + 1)}
  \sum_{m_e m_v}
  |J_P|^2
& = &
  \frac{C_{ev}}{3} 
  (2 J_v + 1)
  \frac{\omega^2}{m_e^2} 
  \,,
\\
\frac{1}{(2 J_e + 1)}
\sum_{m_e m_v}
  J_A^\mu J_P^*
& = &
  \frac{2 C_{ev}}{3} 
  (2 J_v + 1)
  \left( 0, \frac {\bf  q} {m_e} \right).
 \label{eq:ps_Sij}
\qquad
\end{eqnarray}
\label{eq:JAmuJP}
\end{subequations}
Since the emitted photon energy or momentum
transfer is much smaller than the electron mass,
$\omega, {\bf q} \ll m_e$, the pseudo-scalar
atomic current introduces a suppression factor
$\omega^2/m_e^2 \sim 10^{-11}$ for the new
non-interference term and $\omega/m_e 
\sim 10^{-5}$ for the interference one, as
indicated by \geqn{eq:JAmuJP}. So the interference
term would dominate. Putting things together, the
averaged matrix element squared
\geqn{eq:MW_average} becomes,
\begin{subequations}
\begin{eqnarray}
  \overline{|\mathcal{M}^{ij}_S|^2}
& = &
  \frac {2 (2 J_v + 1)|y^e|^2 C_{ev}}
  {[E_{eg} (E_{eg} - 2 \omega) - m^2_\phi]^2}
  \frac{\omega^2}{3 m_e^2}
\\ & \times & \nonumber 
  \left\{
    |y^\nu_{S,ij}|^2
    (p_i \cdot p_j - m_i m_j) 
\right.
\\ & &
+
\left.
    (1 - \delta_M)
    |y^\nu_{P,ij}|^2
    (p_i \cdot p_j + m_i m_j) 
  \right\},
\nonumber
\\
\overline{\mathcal M^{ij}_W \mathcal M^{ij *}_S}
& =  &
  \frac {
    2(2 J_v + 1)\sqrt{2} G_F 
    y^e a_{ij} C_{ev}
  }
  {3 m_e[E_{eg} (E_{eg} - 2 \omega) - m^2_\phi]}
\\
 &  & 
\times
  \left\{
    y_{S,ij}^\nu
    \left[ 
       m_i ({\bf p}_j \cdot {\bf q})
    -
      m_j ({\bf p}_i \cdot {\bf q})
    \right]
\right.
\nonumber
\\ & & 
\left.
   i
   (1 - \delta_M)
    y_{P,ij}^\nu 
    \left[ 
       m_i ({\bf p}_j \cdot {\bf q})
    -
      m_j ({\bf p}_i \cdot {\bf q})
    \right]
    \right\},
\nonumber
\end{eqnarray}
\end{subequations}
where the pseudo-scalar part disappears for
the Majorana case. 

Combining everything into the differential 
cross-section in \geqn{eq:diff_xsec} 
and integrate over $E_{\nu}$, we can calculate the
change in the SM spectral function due to the BSM 
scalar and pseudo-scalar interactions,
\begin{eqnarray}
  \mathcal I_\phi
=
\sum_{ij} 
\frac{\Delta_{ij}(\omega) }
{(E_{vg} - \omega)^2} 
\Theta(\omega - \omega_{ij}^{\rm max})
  \left[ 
    I_{ij}^{\rm SM}(\omega)
  +
    \delta I_{ij}(\omega)
  \right],
\qquad
\end{eqnarray}
where the correction term $\delta I_{ij}(\omega)$ is,
\begin{widetext}
\begin{eqnarray}
  \delta I_{ij}
& = &
\frac{|y^e|^2 \omega^2 }{m_e^2 G_{\rm F}^2}
\frac{
\left[
  |y^\nu_{S,ij}|^2 + (1 - \delta_M)|y^\nu_{P,ij}|^2
\right] 
  E_{eg} (E_{eg} - 2 \omega)
- |y^\nu_{S,ij}|^2 (m_i + m_j)^2
-  (1 - \delta_M) |y^\nu_{P,ij}|^2 (m_i - m_j)^2
}{24[E_{eg}(E_{eg}-2\omega) - m^2_\phi]^2}
\label{eq:dI-scalar}
\\
& + & \nonumber
\frac{y^e\omega^2}{6\sqrt{2} G_F} 
\left\{
\frac { 
{\rm Re}\left[a_{ij} y^\nu_{S,ij}\right]
 (m_i - m_j)
\left[ 
  1 
-
  \frac{(m_i + m_j)^2}{E_{eg}(E_{eg} - 2\omega)}
\right]
  }
{m_e [E_{eg} (E_{eg} - 2 \omega) - m^2_\phi]}
-
(1 - \delta_M)
\frac { 
 {\rm Im}\left[a_{ij} y^\nu_{P,ij}\right]
 (m_i + m_j)
\left[ 
  1 
-
  \frac{(m_i - m_j)^2}{E_{eg}(E_{eg} - 2\omega)}
\right]
  }
{m_e [E_{eg} (E_{eg} - 2 \omega) - m^2_\phi]}
\right\}.
\end{eqnarray}
\end{widetext}

In \gfig{fig:new_spectral_function} we show the 
changes in the spectral function $\mathcal I$ 
due to a light scalar/pseudo-scalar mediator with mass
$m_{\phi} = 20$ keV and couplings
(1) $y^e y^\nu_{S,11} = 6\times 10^{-9}$ (blue
dot-dashed) or (2) $y^e y^\nu_{P,11} = 6\times 10^{-9}$
(brown dotted). Both the scalar and pseudo-scalar
contributions have almost the same shape for Dirac neutrinos as shown in \gfig{fig:new_spectral_function}
since the only difference is a sign in 
front of $m_j$ shown in \geqn{eq:dI-scalar} and the mass term is suppressed when compared with the leading energy term. For Majorana neutrinos, only the scalar couplings can contribute a correction.

\section{Sensitivity and Physics Reach}
\label{sec:sensitivity_New_phys}

To estimate the sensitivity of new physics
parameter, we follow the experimental setup proposed
in~\cite{Song:2015xaa} and take a typical experiment
configuration with $V = 100$ cm$^3$ as well as $n_a = n_\gamma 
= 10^{21}$cm$^{-3}$. 
In addition, the experiment will
run for equal time $T$ at three different frequencies $\omega_i = 1.0688, 1.0699, 1.0711$~eV. Although other works \cite{Zhang:2016lqp,Huang:2019phr} propose
different frequency locations, our results presented here should not change much since the effect is not sensitive to the threshold locations but rather event rate.
The QED background $|e\rangle \rightarrow |g\rangle + n\gamma$ with $n \geq 3$
photons in the final state can be large
\cite{Yoshimura:2015fna}. Fortunately, this background
can be removed using the photonic crystals wave
guides \cite{Tanaka:2016wir} that forbid the emission of photons with a wavelength $\lambda \leq \pi/L$, where
$L$ is the size of the wave guide. Recent advances show that the background can be significantly reduced to achieve background free environment \cite{Tanaka:2019blr}.

The sensitivity is evaluated with the $\chi^2$ function
below,
\begin{eqnarray}
  \chi^2
\equiv
  2 \sum_{\omega_i}
\Bigl\{
&&
  N^{\rm true}(\omega_i) - N^{\rm test}(\omega_i)
\nonumber
\\
- & &
  N^{\rm true}(\omega_i)
  \log \left[ N^{\rm true}(\omega_i)/N^{\rm test}(\omega_i) \right]
\Bigr\},
\quad
\end{eqnarray}
in terms of the total numbers
$N^{\rm true}(\omega_i)$ and $N^{\rm test}(
\omega_i)$ of the true and test event  samples,
respectively,
\begin{eqnarray}
  N (\omega)
\equiv
 0.002
  \left(\frac{T}{s}\right) 
  \left(\frac{V}{100~{\rm cm}^3}\right)
  \left(\frac{n_a \mbox{ or } n_\gamma}{10^{21} {\rm cm}^{-3}}\right)^3
  \mathcal I (\omega).
\qquad
\end{eqnarray}
For a test experiment, we expect $\mathcal O(20)$ events
with an exposure time $T = 2.3$ days for the SM
interactions. 

\begin{widetext}
\centering
    \begin{minipage}{0.99\linewidth}
\begin{figure}[H]
\centering
\includegraphics[scale = 0.23]{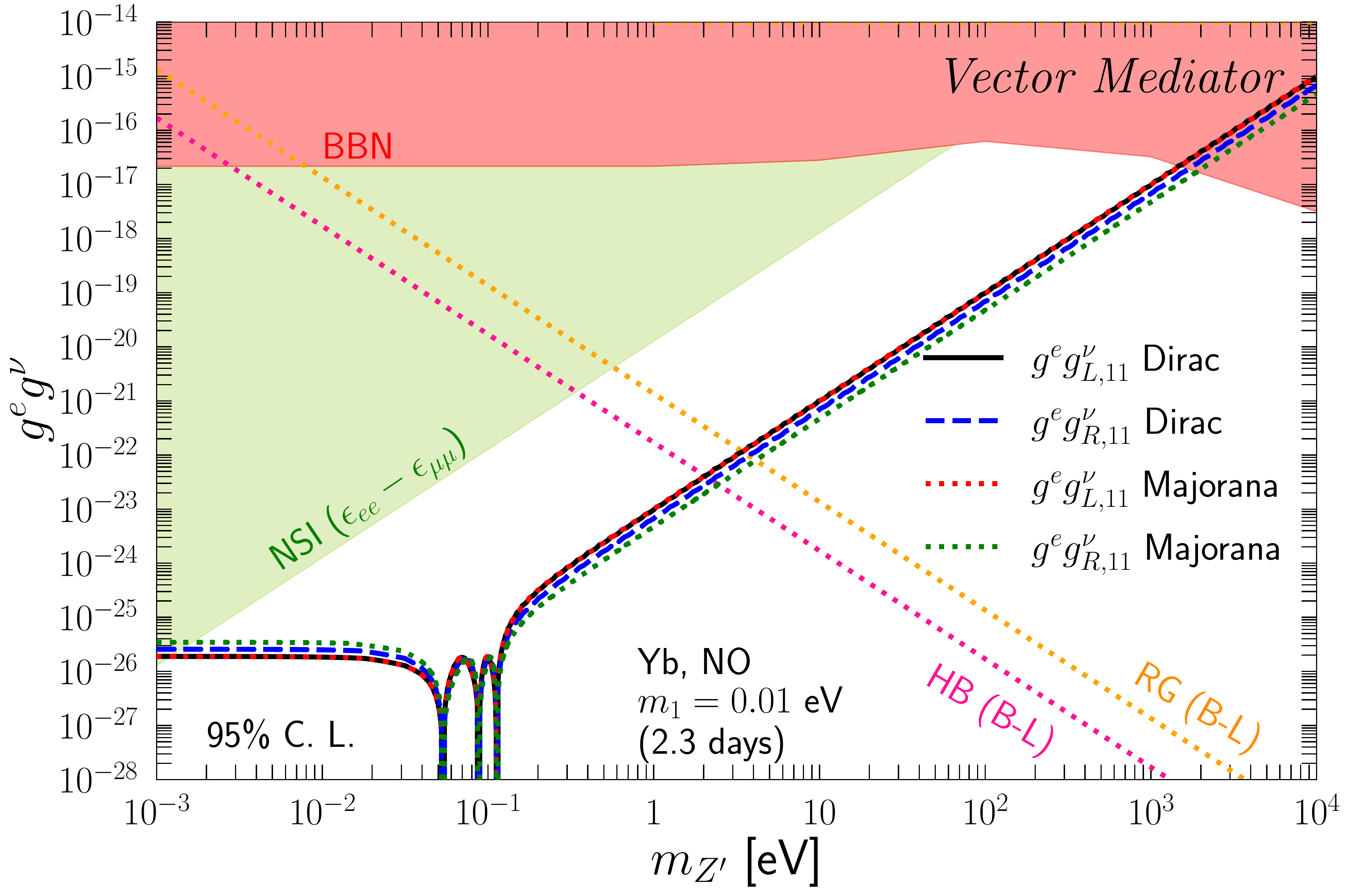}
\hfill
\includegraphics[scale = 0.23]{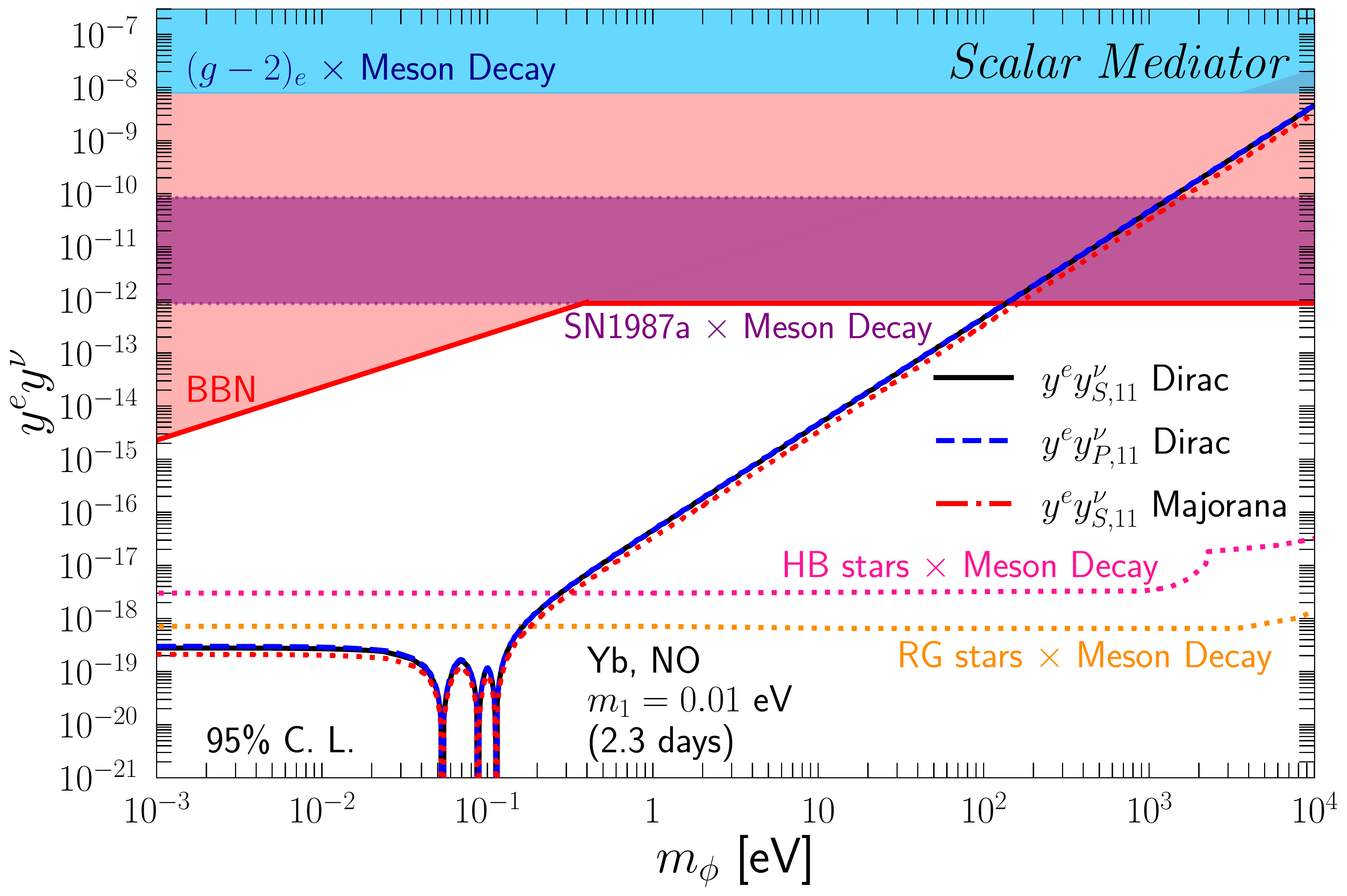}
\caption{{\bf Left:} 
Sensitivity on the combinations $\left|g^e
g^\nu_{L,ij}\right|$ and $\left|g^e
g^\nu_{R,ij}\right|$ as a function of the vector mediator
mass $m_{Z'}$ for Dirac and Majorana neutrinos.
{\bf Right:} 
Sensitivity on the combinations $|y^e y^\nu_{S,ij}|$,
and $|y^e y^\nu_{P,ij}|$ as a function of the scalar
mediator mass $m_{\phi}$. In these plots
we assume the normal neutrino mass ordering (NO) with the lightest mass $m_1 = 0.01$ eV and an exposure time of $T = 2.3$ days at a Yb-based experiment.}
\label{fig:sensitivity}
\end{figure}
    \end{minipage}
\end{widetext}
The sensitivity curves shown in
\gfig{fig:sensitivity} are obtained for $\Delta
\chi^2 = 5.99$,  corresponding to 95\%
C.L. The region above these curves will be excluded if no signal observed. We can see that the sensitivity
on the coupling constants decreases with increasing mediator
mass for both vector and scalar cases.
The sensitivity can reach $\left|g^e g^\nu_{L,ij}\right|$, $\left|g^e
g^\nu_{R,ij}\right| \lesssim 10^{-15} \sim 10^{-26}$
($|y^e y^\nu_{S,ij}|$,
$|y^e y^\nu_{P,ij}| \lesssim 10^{-9} \sim 10^{-19}$)
for the vector (scalar) mediator. In both cases, we take one
diagonal neutrino coupling in the mass basis,
$g_{11}, y_{11} \neq 0$, at a time. Notice that there
is no $y^\nu_P$ curve for Majorana
neutrinos since the pseudo-scalar
term va\-ni\-shes in this situation, as
noted in \geqn{eq:dI-scalar}.

The behaviour of the curve can be understood as follows. For large mass, $m_{Z',\phi}$ dominates
over $E_{eg}(E_{eg}-2\omega)$ and the curve
decreases with $m_{Z',\phi}^2$ as a line with
slope $2$ in the log-log scale. For very small
mass, the situation is the opposite and $E_{eg}
(E_{eg}-2\omega)$ dominates over $m_{Z',\phi}$. As
result, the event rate becomes insensitive to the
mediator mass and the sensitivity curve flattens.
In between, the propagator has a pole at $\omega =
\frac{E_{eg}}{2}- \frac{m_{Z',\phi}^2}{2E_{eg}}$
and hence the diverging sensitivity. 
Since our benchmark experiment has only three
frequency points, $\omega = 1.0688, 1.0699,
1.0711$\,eV, the divergence happens at three
masses $m_{Z',\phi} = 0.11, 0.088, 0.053$\,eV,
respectively. However, these dips in the
sensitivity curves are not physical. The mediator
decay into a pair of neutrinos would lead
to a nonzero decay width $\Gamma_{Z', \phi}$ in
the mediator propagator and hence modulate the
sensitivity behavior. But it is not
straightforward to implement in
\gfig{fig:sensitivity} since the decay
width only relies on the neutrino coupling $g^\nu$
or $y^\nu$. For a fixed value of the vertical axis
$g^e g^\nu$ ($y^e y^\nu$), there are too many
choices for $g^\nu$ ($y^\nu$). So there is no
unique way of implementing the mediator decay width in
\gfig{fig:sensitivity}. For simplicity,
the dips are kept but we shall keep in mind that
the sensitivity is not that dramatic at these
points.

\gfig{fig:sensitivity} also shows
that there is no difference in the sensitivity of
$g^e g_{L,11}^\nu $ between Dirac and
Majorana neutrinos. This happens because the extra
contribution due to the Majorana property,
$I^{(M)}_{ij}$ of \geqn{eq:I_Dirac_and_Maj}, is small.
Taking $m_1 = 0.01$\,eV,
$I^{(M)}_{ij}/I^{(D)}_{ij} \sim 6m_i m_j /
\omega^2 \approx 10^{-3}$ which is negligible.
When only $g_{L}^\nu$ coupling is non-zero, the change
to the $\mathcal I_{Z'}$ in \geqn{eq:Iz} that is proportional
to $I^{(M)}$ is tiny. For 
$g_{R}^\nu \neq 0$, the situation is different.
There is a mixing term contribution for Majorana
neutrinos, $-2{\rm Re}[a^L_{ij} a^R_{ij}]
I^{(D)}_{ij}$ in \geqn{eq:Iz} that can be large.

For comparison, also show several other constraints
in \gfig{fig:sensitivity}. The vector mediator case has
three bounds: The BBN bound for a $Z'$ was obtained from the relativistic degrees of freedom
$N_{\rm eff}$ in the early Universe \cite{Ibe:2020dly}.
Since BBN happens at $\mathcal O(\mbox{MeV})$, the light
mediator mass in the considered range $[10^{-3}, 10^4]$\,eV
is negligibly small. Consequently, the BBN bound is insensitive
to the $Z'$ mass and flat. The red giant (RG) and horizontal
branch (HB) bounds from stellar cooling \cite{Hardy:2016kme}
are based on a specific $B-L$ model with the coupling to leptons.
The dotted curves for the RG and HB constraints with
$U(1)_{B-L}$ in the left panel of \gfig{fig:sensitivity} are then obtained 
from the bounds on $g_{B - L}$ and the model property that $g^e  = g^\nu = g_{B- L}$.
Finally, the NSI bound (green) is obtained from the global fit of neutrino experiments \cite{Coloma:2020gfv}.
Since the NSI effect is due to the coherent scattering of neutrinos in matter with zero momentum transfer, it is only sensitive to the ratio between coupling and
mediator mass, $g^e g^\nu / m^2_{Z'}$, reflected as a straight line with slope 2 in the plot.

Most bounds for the scalar mediator case are for the coupling with neutrino or electron separately.
The bound on the pseudo-scalar coupling with the
electron neutrino $\nu_e$, $|y^\nu_{ee}| <
1.5\times 10^{-3}$, is from pion 
\begin{widetext}
\centering
    \begin{minipage}{0.99\linewidth}
\begin{figure}[H]
\includegraphics[scale = 0.23]{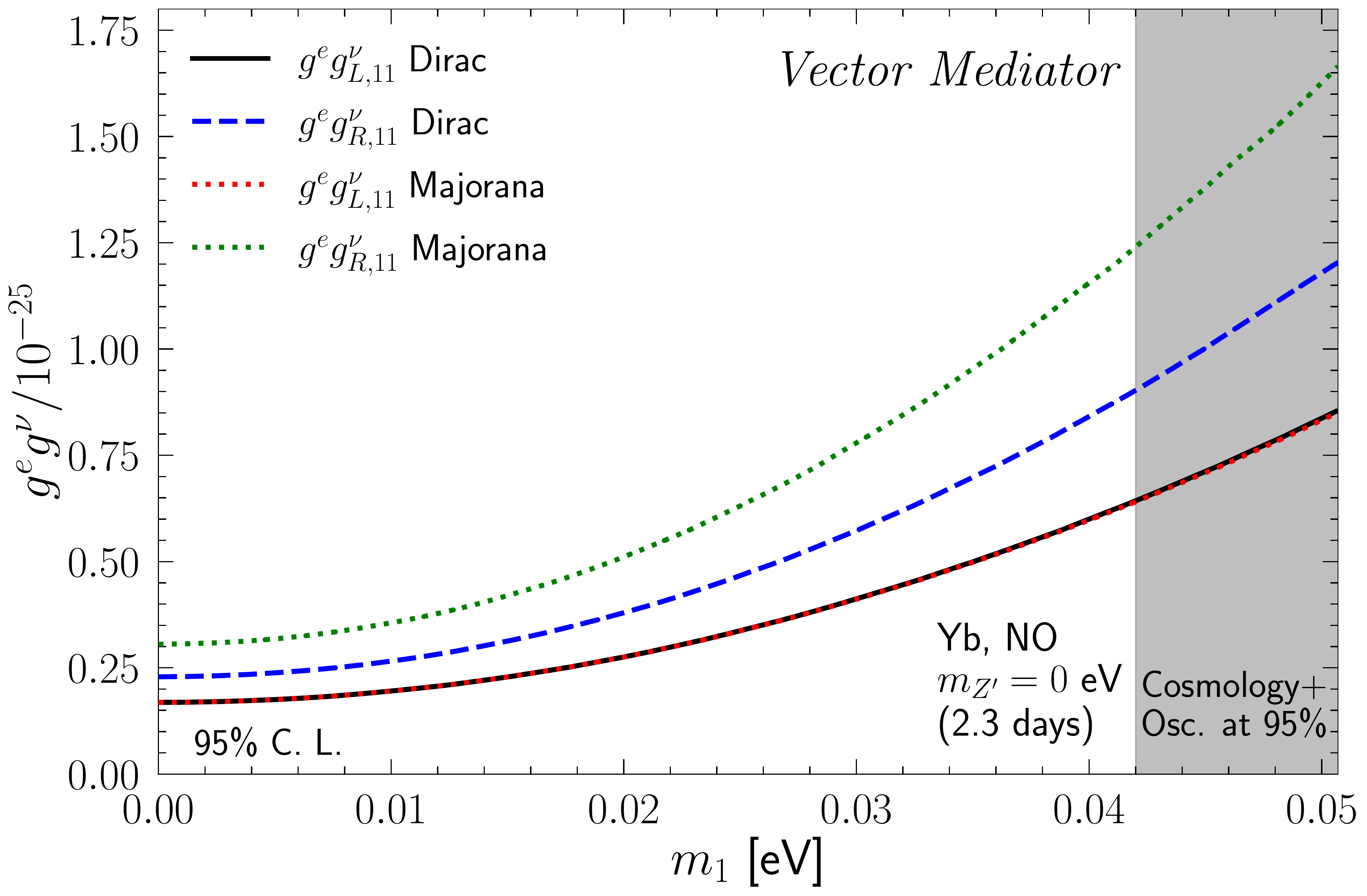}
\hfill
\includegraphics[scale = 0.23]{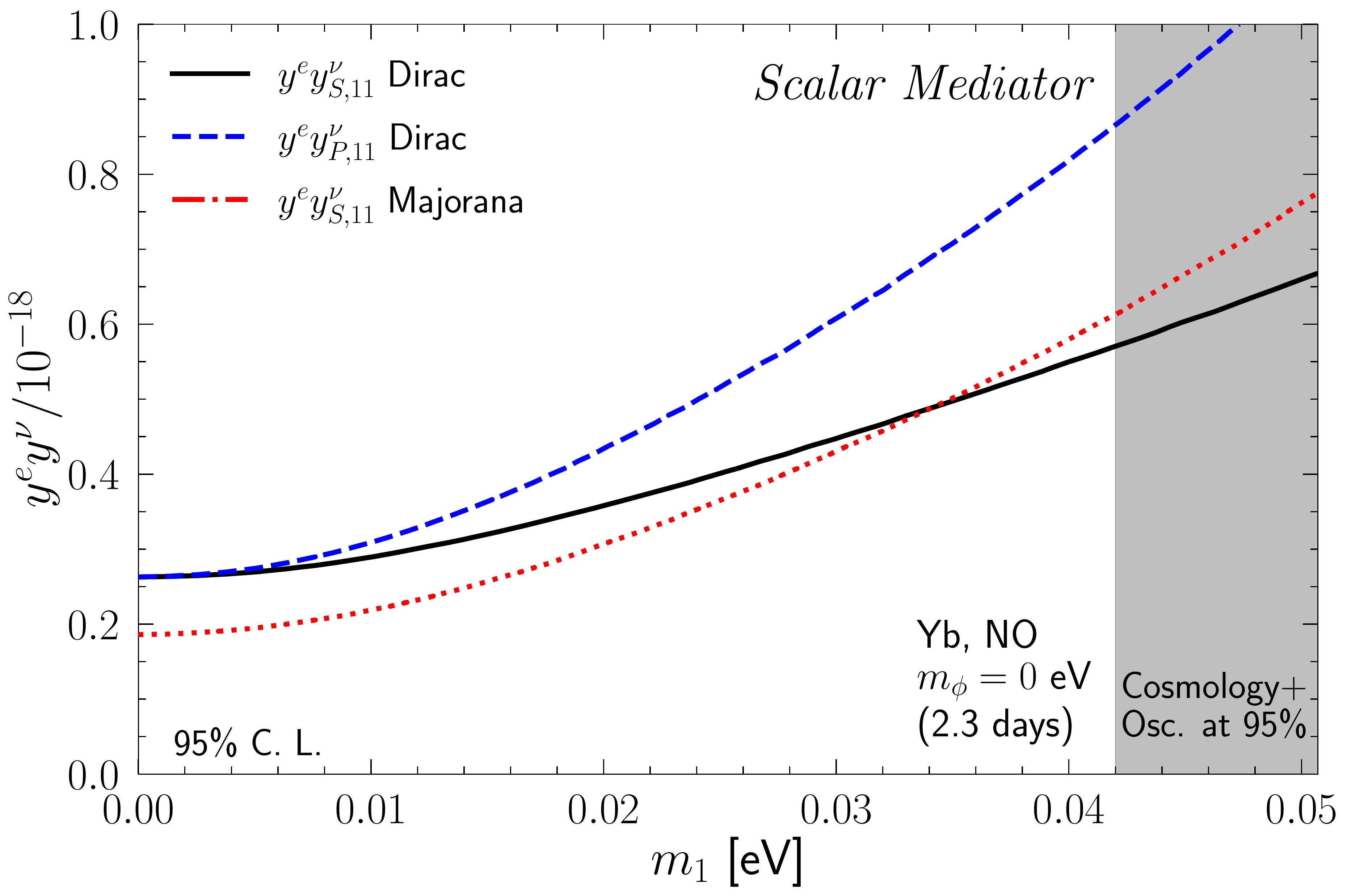}
\caption{
Sensitivity on the coupling combinations $\left|g^e
g^\nu \right|$ for vector mediator (left) and $\left|y^e
y^\nu \right|$ for the scalar one (right) as a function
of the lightest neutrino mass $m_1$ with the normal mass
ordering. For both cases, a vanishing mediator
mass $m_{Z'}, m_\phi = 0$\,eV is adopted. To make comparison,
we have considered both Dirac and Majorana neutrinos.
These sensitivities are obtained with an exposure of
$T = 2.3$\,days at a Yb-based experiment. The gray region
shows the combined constraint from the current cosmological
data \cite{Zyla:2020zbs} and neutrino oscillation
measurements \cite{deSalas:2020pgw}.}
\label{fig:sensitivity_vrs_mnu}
\end{figure}
    \end{minipage}
\end{widetext}
\noindent 
decay
\cite{Pasquini:2015fjv,Berryman:2018ogk,deGouvea:2019qaz,Dror:2020fbh}.
For the coupling with
electron, the bound comes from the electron
anomalous magnetic moment $(g-2)_e$ measurement
\cite{Mohr:2015ccw} which is the most stringent
experiment on Earth. Astrophysical sources like RG
and HB can also provide constraints from the star
cooling process \cite{Knapen:2017xzo, DeRocco:2020xdt}. If these bounds follow a
half-normal distribution, the pro\-du\-ct
distribution follows the Bessel function of
the first kind and the 95\% upper bound is a
product of individual ones, $(y^e y^\nu)_{95\%} \approx
0.57 \times (y^e)_{95\%} \times (y^\nu)_{95\%}$,
to combine the separate bounds on electron and
neutrino couplings. 

We also include a BBN bound
\cite{Venzor:2020ova} due to the effective
neutrino mass via matter effect \cite{Ge:2018uhz}
in the early Universe. This is the only bound
implementing the electron and neutrino couplings
simultaneously. We also include a more general bound arising
from the production of relativistic degrees of
freedom. The result in \cite{DeRocco:2020xdt} was obtained 
for the electron coupling only,
$y^e < 10^{-9}$. In order to obtain the
bound for the combination $y^ey^\nu$, we combine 
with the pion decay data as before, resulting in
$\left(y^e y^\nu \right)_{95\%} < 10^{-12}$.
This bound starts to dominate the BBN constraint
at $m_\phi > 0.4$\,eV and is of same size as the
lower region of the SN bound.

With coupling to electron, a light mediator, such
as axion \cite{Huang:2019rmc} and dark photon
\cite{Bhoonah:2019eyo}, can also be directly
produced in this coherently stimulated
process, $|e\rangle \rightarrow |g\rangle + \gamma
+ a/\gamma'$. Experimentally, there is no
distinction whether the final state is a neutrino
pair, axion, or dark photon. Although it can also
provide a sensitive probe, this direct production
mode can only test the coupling with electron but
not the one with neutrinos. In addition, the
emitted axion or dark photon cannot be heavier
than the emitted energy $E_{eg}$. For comparison,
the neutrino pair emission with light mediator as
an intermediate particle can simultaneously probe
the couplings to both electron and neutrino
without limitation on the mediator mass.

\gfig{fig:sensitivity_vrs_mnu} shows the dependence of
the coupling sensitivities on the lightest neutrino mass
$m_1$. To avoid complication from the mediator mass
as shown in \gfig{fig:sensitivity}, we focus on the massless
mediator limit where the sensitivity is a constant and
essentially independent of $m_{Z'}/m_\phi$. The normal
ordering is adopted for illustration with the corresponding
lightest mass $m_1$ varies from 0\,eV to 0.05\,eV.
The neutrino pair emission is more sensitive with vanishing
$m_1$. Increasing $m_1$ to 0.05\,eV introduces roughly
a factor of 4 reduction in the sensitivity.
This happens because the allowed phase-space shrinks
with increasing neutrino mass and consequently smaller
event rates. Again, there is no difference in
the sensitivity for $g^e g_{L,11}^\nu$ between Dirac
(black line) and Majorana (dotted red line) neutrinos
as shown in the left panel of \gfig{fig:sensitivity_vrs_mnu}.
This happens because contribution $I_{ij}^{(M)}$ purely
due to the Majorana property only contributes $10^{-3}$
of $\mathcal I_{Z'}$ in \geqn{eq:Iz}.
For comparison, the gray region shows the 
95\% C.L. upper bound on the lightest neutrino mass $m_1$
from cosmology \cite{Zyla:2020zbs} and oscillation
measurements \cite{deSalas:2020pgw}.

\section{Conclusion}
\label{sec:conclusion}

The sensitivity to light mediators is typically
limited by the involved momentum transfer. The
various existing constraints on the coupling
constant can no longer improve once the light
mediator mass decreases to become comparable with
momentum transfer. This is most transparent in the
BBN, RG, and HB constraints in
\gfig{fig:sensitivity}. The only
exception is the NSI bound with neutrino coherent
scattering but its sensitivity is limited to the
neutrino oscillation experiment precision which is just
entering percentage era and it cannot disentangle
coupling from the mediator mass. In this paper, we
point out the promising future of the neutrino pair
emission process as a very sensitive probe of light
vector and scalar mediators around the $\mathcal O$(eV)
scale. It can enhance the sensitivity by 3 orders more
than the NSI above $\mathcal O(0.1$eV) for the
vector mediator case and the BBN bound for the
scalar case. The best upper limit can be as low as
$|y^e y^\nu| < 10^{-9} \sim 10^{-19}$ for a scalar
mediator and $|g^e g^\nu| < 10^{-15} \sim
10^{-26}$ for a vector one, respectively. The RENP
process can not only become an ideal place for
testing the neutrino nature and parameters but
also provide a sensitive probe of the associated
new physics with neutrinos.

\section*{Acknowledgements}

This work is supported by the Double First Class start-up fund
(WF220442604), the Shanghai Pujiang Program (20PJ1407800),
National Natural Science Foundation of China (No. 12090064),
and Chinese Academy of
Sciences Center for Excellence in Particle Physics (CCEPP).
The authors are grateful to Xiao-Dong Ma and Jin Sun
for reading the draft and providing useful comments.

\end{document}